\documentclass[journal,onecolumn,draftclsnofoot]{IEEEtran}
\ifCLASSINFOpdf
\else
\fi
\usepackage[utf8]{inputenc} 
\usepackage[T1]{fontenc}    
\usepackage{hyperref}       
\usepackage{url}            
\usepackage{booktabs}       
\usepackage{amsfonts}       
\usepackage{nicefrac}       
\usepackage{microtype}      

\usepackage{dsfont}
\usepackage[cmex10]{amsmath}
\usepackage{amsfonts,amssymb,amssymb,mathrsfs,mathdots,mathtools}
\makeatletter
  \def\my@tag@font{\normalsize}
  \def\maketag@@@#1{\hbox{\m@th\normalfont\my@tag@font#1}}
  \let\amsmath@eqref\eqref
  \renewcommand\eqref[1]{{\let\my@tag@font\relax\amsmath@eqref{#1}}}
\makeatother

\usepackage{xcolor}
\usepackage{graphicx}
\usepackage{float}
\usepackage{color, soul}

\usepackage{stmaryrd}
\usepackage{verbatim}
\usepackage{paralist}
\usepackage{dirtytalk}
\usepackage{flushend}

\pdfoutput=1
\newtheorem{theo}{Theorem}
\newtheorem{prop}[theo]{Proposition}
\newtheorem{lem}[theo]{Lemma}
\newtheorem{cor}[theo]{Corollary}
\newtheorem{rem}[theo]{Remark}
\newtheorem{defi}[theo]{Definition}
\newtheorem{exe}[theo]{Example}

\makeatletter
\renewcommand*\env@matrix[1][*\c@MaxMatrixCols c]{%
  \hskip -\arraycolsep
  \let\@ifnextchar\new@ifnextchar
  \array{#1}}
\makeatother

\DeclarePairedDelimiter\abs{\lvert}{\rvert}

\definecolor{ballblue}{rgb}{0.13, 0.67, 0.8} 

\DeclareMathOperator*{\argmax}{arg\,max}

\graphicspath{ {./figures/} }

\newcounter{relctr} 
\everydisplay\expandafter{\the\everydisplay\setcounter{relctr}{0}} 

\newcommand\labelrel[2]{%
  \begingroup
    \refstepcounter{relctr}%
    \stackrel{\textnormal{(\alph{relctr})}}{\mathstrut{#1}}%
    \originallabel{#2}%
  \endgroup
}
\AtBeginDocument{\let\originallabel\label} 

\begin{document}
%
\title{Quantifying Membership Privacy via Information Leakage}
%
%
%

\author{Sara~Saeidian,~\IEEEmembership{Student~Member,~IEEE,}
        Giulia~Cervia,~\IEEEmembership{Member,~IEEE,}
        Tobias~J.~Oechtering,~\IEEEmembership{Senior~Member,~IEEE} 
        Mikael~Skoglund,~\IEEEmembership{Fellow,~IEEE}
}

\maketitle

\begin{abstract}
Machine learning models are known to memorize the unique properties of individual data points in a training set. This memorization capability can be exploited by several types of attacks to infer information about the training data, most notably, membership inference attacks. In this paper, we propose an approach based on information leakage for guaranteeing membership privacy. Specifically, we propose to use a conditional form of the notion of \textit{maximal leakage} to quantify the information leaking about \textit{individual} data entries in a dataset, i.e., the entrywise information leakage. We apply our privacy analysis to the \textit{Private Aggregation of Teacher Ensembles} (PATE) framework for privacy-preserving classification of sensitive data and prove that the entrywise information leakage of its aggregation mechanism is Schur-concave when the injected noise has a log-concave probability density. The Schur-concavity of this leakage implies that increased consensus among teachers in labeling a query reduces its associated privacy cost. Finally, we derive upper bounds on the entrywise information leakage when the aggregation mechanism uses Laplace distributed noise.
\end{abstract}

%

%
\IEEEpeerreviewmaketitle

\section{Introduction}
\label{sec:intro}
In recent years, many useful machine learning applications have emerged that require training on sensitive data. Such applications span across a diverse range of fields such as medical imaging~\cite{litjens2017survey}, rumor identification in social media~\cite{liang2015rumor}, or financial fraud detection~\cite{west2016intelligent}. While all machine learning applications by definition reveal some information about the training data, privacy concerns arise when machine learning models memorize properties that are \textit{unique} to individual data entries. In fact, a variety of privacy attacks have demonstrated that it is indeed possible to exploit this \say{memorization} capability of models to infer information about data entries in the training set~\cite{papernot2018sok}. 

Arguably, the simplest type of privacy attacks against machine learning models is \textit{membership inference} attacks in which an adversary infers whether or not a certain data point was used in the training~\cite{shokri2017membership, long2018understanding}. In response to such attacks, a number of mitigation techniques have been proposed in the literature, with \textit{differential privacy}-based methods being the most commonly studied. Differential privacy~\cite{dwork2014algorithmic} provides provable and operationally meaningful privacy guarantees, and by definition neutralizes membership inference attacks. Roughly speaking, differential privacy ensures that all datasets differing in only one entry (i.e., adjacent datasets) produce an output with similar probabilities. Moreover, it has several useful properties, such as satisfying data-processing inequalities and composition theorems~\cite{dwork2014algorithmic}. 

The standard definition of differential privacy (i.e., pure differential privacy) uses a parameter $\epsilon$ to define a multiplicative upper bound on the changes in the probability of an output for all adjacent datasets in the input~\cite{dwork2006calibrating}. However, this definition is known to be very strict, and has limited applicability. As such, several relaxations of differential privacy have been proposed, the most notable of which is $(\epsilon, \delta)$-differential privacy~\cite{dwork2006our}. A common interpretation of $(\epsilon, \delta)$-differential privacy is that the guarantees of $\epsilon$-differential privacy hold except with probability $\delta$. Thus, it provides the necessary flexibility for studying a larger class of privacy-preserving mechanisms such as the Gaussian mechanism~\cite{dwork2006calibrating}.  

Despite the advantages of $(\epsilon,\delta)$-differential privacy, one should note that its privacy guarantees are qualitatively different from those of pure differential privacy (see~\cite{mironov2017renyi} for illustrative examples). On this account, recently R{\'e}nyi differential privacy~\cite{mironov2017renyi} was proposed as an alternative relaxation of pure differential privacy. While R{\'e}nyi differential privacy satisfies the same useful properties as pure differential privacy, it does not offer any intuitive operational meaning, and its privacy guarantees are usually translated into $(\epsilon, \delta)$-differential privacy for interpretation.
  
In this paper, we propose to use (a conditional form of) the notion of \textit{maximal leakage}~\cite{issa2019operational} to measure the amount of information leaking about any single data entry in a dataset, i.e., the entrywise information leakage. Maximal leakage~\cite{issa2019operational} is an operationally meaningful privacy metric that captures the inference capabilities of an adversary trying to deduce some information about the input data by observing the output. Specifically, maximal leakage \textit{quantifies} the maximal gain in an adversary's ability to correctly guess any arbitrary discrete function of the input data by observing the output (as opposed to making a guess with no observations). Note that the original definition of maximal leakage quantifies the information leaking about the \textit{whole dataset}, whereas we are interested in measuring the information leaking about \textit{single data entries} in the dataset. As such, similarly to~\cite{alvim2011differential}, we consider an adversary who knows the values of all the entries in the dataset, except for a single data entry of interest. Intuitively, in this setup, observations only convey the \textit{unique} information contributed by the unknown data entry since all other entries are already known to the adversary. To quantify this entrywise information leakage, we propose a conditional form of maximal leakage, namely the \textit{pointwise conditional maximal leakage}, which is also a special case of the event-conditional Sibson mutual information introduced in~\cite{liao2019robustness}. Then, by allowing the unknown entry to be any of the entries in the dataset, we can derive upper bounds on the entrywise information leakage, and provide meaningful worst-case privacy guarantees. 

Maximal leakage satisfies several useful properties, most notably a data-processing inequality and a composition lemma~\cite{issa2019operational}. The data-processing inequality ensures that no manipulation of the output can increase the information leakage, while the composition property characterizes the information leaked through multiple observations. Here, we show that the same properties hold for pointwise conditional maximal leakage, rendering it suitable for privacy analysis of more complex information systems.

We apply our privacy analysis to the \textit{Private Aggregation of Teacher Ensembles} (PATE) framework~\cite{papernot2016semi, papernot2018scalable}. PATE is a general framework for privacy-preserving classification of sensitive data, and operates by transferring the knowledge of an ensemble of models (called \textit{teachers}) trained on disjoint partitions of the sensitive data to a \textit{student} classifier. Specifically, the student is trained using a public unlabelled dataset which will be labelled by the teachers through an \textit{aggregation mechanism}. The aggregation mechanism is essentially the \textit{Report-Noisy-Max mechanism}~\cite{dwork2014algorithmic} which adds noise to the teachers' predictions to enable derivation of privacy guarantees.

PATE has several advantages as a privacy-preserving machine learning framework. First, the privacy guarantees result solely from the aggregation mechanism and are agnostic to the specific machine learning techniques used by each teacher. This is because the modular structure of PATE enables us to invoke the data-processing inequality to uncouple the information leaked through the training and aggregation, and guarantee that the overall leakage is less than both. Second, PATE lends itself well to distributed learning by allowing data owners to separately train their own predictors, hence mitigating the need for centralized storage of the sensitive data. Finally, the aggregation mechanism induces a favorable \textit{synergy between privacy and accuracy} such that increased agreement among the teachers in labelling a query lowers its associated privacy cost. This synergy is one of the main focuses of this paper, and will be extensively studied.

The privacy guarantees established by PATE are characterized in~\cite{papernot2016semi,papernot2018scalable} in terms of differential privacy, and results from experiments are reported. However, these works do not analytically prove the aforementioned synergy between privacy and accuracy observed in the framework. Here, we will analyze the privacy of the framework in terms of the entrywise information leakage, and prove the privacy-accuracy synergy using analytical arguments in order to provide deeper insights into the workings of the framework, especially the Report-Noisy-Max mechanism used for aggregating teachers' predictions. As~\cite{papernot2016semi,papernot2018scalable} present a thorough experimental study, here we refrain from repeating the experiments but focus on giving a rigorous theoretical analysis of the framework.

\subsection{Contributions}
Our contributions can be summarized as follows:
\begin{enumerate}[i)]
\item \textbf{Introducing pointwise conditional maximal leakage.} We approach membership privacy from a novel angle by studying the information leakage of individual data entries in a database. We begin by deriving a data-processing inequality and a composition lemma for pointwise conditional maximal leakage, and then apply them to the problem of studying the entrywise information leakage in PATE. 

\item \textbf{Proving the privacy-accuracy synergy in PATE.} We show that the entrywise information leakage of the aggregation mechanism in PATE (i.e., the Report-Noisy-Max mechanism) is \textit{Schur-concave}~\cite{marshall1979inequalities, jorswieck2007majorization} when the injected noise has a \textit{log-concave}~\cite{bagnoli2005log, an1997log} probability density. As we will see, this implies that increased consensus among teachers lowers the privacy cost of labelling a query. Note that many commonly used probability distributions including the Laplace and Gaussian distributions are log-concave rendering this result fairly general.

\item \textbf{Deriving membership privacy guarantees for PATE with Laplace noise.} We derive upper bounds on the entrywise information leakage when the noise injected in the aggregation mechanism has Laplace distribution. We present two types of bounds: a data-independent bound, which holds uniformly for all training datasets and is tight in the sense that the bound holds with equality when the information leakage is maximized. Our other bound is data-dependent in that it depends on the training data through the teachers' predictions. The data-dependent bound can be tighter than the data-independent bound when there is a large consensus among the teachers in predicting the label of a query.
\end{enumerate}

\subsection{Other Related Work}
\textbf{Information leakage metrics.} In recent years, a large body of work has been dedicated to studying various information-theoretic privacy metrics. Most notably, mutual information has been frequently proposed and studied as such a metric (see e.g., \cite{prabhakaran2007secure, sankar2013utility, wang2016relation}) by appealing to its operational meaning in communication theory. Similarly, in~\cite{rassouli2020optimal} another information-theoretic quantity namely the total variation distance is studied as a privacy metric in an information disclosure scenario. More closely related to our approach, several information leakage metrics have recently emerged that aim to capture the inference abilities of an adversary trying to guess a secret. For instance, \cite{asoodeh2017privacy} proposes to use the probability of correctly guessing the secret as a privacy metric. In~\cite{liao2019tunable} a class of tunable loss functions are introduced to capture a range of adversarial objectives, e.g., refining a belief or guessing the most likely value for the secret. Other methods include posing the privacy problem as a hypothesis test, e.g., in~\cite{li2019privacy}. It is worth mentioning that some of the proposed privacy metrics (such as mutual information and total variation distance) have no clear operational meaning in the privacy setting, which limits their applicability. A systematic survey of privacy metrics is provided in~\cite{wagner2018technical}.  

\textbf{Privacy-preserving machine learning.} Several centralized and decentralized solutions have been proposed in the literature that provide privacy guarantees in terms of differential privacy. To give a few examples, \cite{shokri2015privacy} proposes a collaborative framework for privacy-preserving deep learning where the guarantees of differential privacy are obtained by perturbing the gradients. Another example is~\cite{abadi2016deep} where the privacy analysis of gradient perturbations are improved by introducing the \textit{moments accountant} framework. Other methods include privacy-preserving logistic regression~\cite{chaudhuri2009privacy, zhang2012functional}, support vector machines~\cite{rubinstein2009learning} and empirical risk minimization~\cite{chaudhuri2011differentially, bassily2014private}.      

\subsection{Outline of the Paper}
The rest of the paper is organized as follows: in Section~\ref{sec:background} we will review the definition of maximal leakage and give a short summary of the operation of the PATE framework. In Section~\ref{sec:pcml_properties} we will present the definition of pointwise conditional maximal leakage, and state a few of its key properties. In Section~\ref{sec:leakage} we will present our privacy analysis of the framework and state our results. Section~\ref{sec:conclusions} concludes the paper.

\section{Background}
\label{sec:background}
Throughout the paper, upper-case letters are used to represent discrete random variables, upper-case calligraphic letters represent their corresponding alphabets and lower-case letters represent the elements of the alphabets. We will use $\llbracket 1,n\rrbracket= \{1, \ldots, n\}$ to denote the set of integers between one and $n$. Let $A = (A_1, \ldots, A_n)$ be a sequence of $n$ elements. We will use the notation $A \setminus A_j$ to denote the sequence of $n-1$ elements obtained by removing the $j$th element in $A$ for some $j \in \llbracket 1,n\rrbracket$. Furthermore, we will use $\abs{\;\cdot\;}$ to denote the cardinality of a set, and $\log(\cdot)$ to denote the natural logarithm. Finally, all sets considered in this paper are assumed to be finite.

We begin by reviewing a few key concepts. 
\subsection{Maximal Leakage}
Let $X$ be a random variable representing the data containing sensitive information, and $Y$ be the publicly observed output of a probability kernel $P_{Y \mid X}$ with input $X$. Suppose that an adversary observes $Y$ and wishes to guess an arbitrary discrete function of $X$, denoted by $U$.     
\begin{defi}[Maximal leakage~\cite{issa2019operational}]
Suppose $P_{XY}$ is a joint distribution defined on the alphabets $\mathcal{X}$ and $\mathcal{Y}$. The maximal leakage from $X$ to $Y$ is defined as 
\begin{equation}\label{eq:def_ml}
\mathcal{L}(X \to Y) \coloneqq \sup_{U:\,U-X-Y} \log \frac{{\mathbb{P}} \left(U = \hat{U}(Y)\right)}{\max_{u \in \mathcal{U}} P_U(u)},
\end{equation}
where $\hat{U}$ is the optimal estimator (i.e., MAP estimator) taking values from the same alphabet as $U$. 
\end{defi}
Maximal leakage quantifies the maximal gain in the adversary's ability to correctly guess $U$ after observing $Y$ (compared to correctly guessing $U$ with no observations). It is shown in~\cite[Theorem 1]{issa2019operational} that for finite alphabets $\mathcal{X}$ and $\mathcal{Y}$, ~\eqref{eq:def_ml} simplifies to
\begin{equation}
\mathcal{L}(X \to Y) = \log \sum_{y \in \mathcal{Y}} \max_{x \in \mathcal{X} : P_X(x)>0} P_{Y \mid X} (y \mid x). 
\end{equation}

\subsection{The PATE Framework}
PATE~\cite{papernot2016semi, papernot2018scalable} is a general framework for privacy-preserving classification of sensitive data. It operates by transferring the knowledge of an ensemble of classifiers, called \textit{teachers}, trained on (disjoint) partitions of the sensitive data to a \textit{student} classifier. More specifically, the PATE framework consists of the following three main components:

\textbf{Teacher models}. A teacher is a classification model trained on one of the disjoint partitions of the sensitive training data, and can use any classification algorithm suited for the task. At inference, each teacher predicts a label independently of others, to which we will refer as that teacher's \textit{vote}. Thus, partitioning data into $L$ sets (and correspondingly training $L$ teachers) produces $L$ primary votes for predicting the label of any new data point.

\textbf{Aggregation mechanism}. To predict the label of a new data point, the aggregation mechanism (i.e., the Report-Noisy-Max mechanism~\cite{dwork2014algorithmic}) constructs the histogram of teachers' votes, adds calibrated noise to each of the bins, and outputs the class label with the maximum noisy vote as the final aggregate prediction. Note that the overall privacy guarantees of the framework result from the addition of noise in the aggregation mechanism. 

\textbf{Student model}.
The student model is trained using a public unlabelled dataset which will be labelled by the teachers' ensemble through the aggregation mechanism. Note that to limit the privacy cost of the overall system, the student must be trained with as few queries to the teachers as possible.

\section{Pointwise conditional maximal leakage}
\label{sec:pcml_properties}
In this section, we introduce the notion of pointwise conditional maximal leakage, and state two of its important properties. Recall that maximal leakage is defined in a setup where an adversary wishes to guess an arbitrary discrete function $U$ of the private input data $X$ by observing the output $Y$. Here, we consider the case where the adversary has some \textit{a priori} knowledge about $X$. We model this a priori knowledge as the outcome of a random variable, and accordingly define a \textit{conditional} form of maximal leakage. Consider an adversary that knows the outcome of a random variable $Z$. 
\begin{defi}[Pointwise conditional maximal leakage]
Suppose $P_{XYZ}$ is a joint distribution defined on the alphabets $\mathcal{X}$, $\mathcal{Y}$ and $\mathcal{Z}$, and that the value of the random variable $Z$ is a priori given as $z \in \mathcal{Z}$. The pointwise conditional maximal leakage from $X$ to $Y$ given $Z=z$ is defined as  
\begin{equation}
{\mathcal L}(X\!\to \!Y|Z=z)  \coloneqq \!\!\!\! \sup_{U: \, U-(X,Z)-Y} \!\! \log{\!\frac{\mathbb{P} \left(U\!= \!\hat U(Y,Z\!=\!z)\right)}{\mathbb{P} \left(U= \tilde U(Z=z)\right)}},
\label{eq:cml1}
\end{equation}
where $\hat{U}$ is the optimal estimator of $U$ given $Y$ and $Z=z$, and $\tilde{U}$ is optimal estimator of $U$ given only $Z=z$. 
\end{defi}

\begin{prop}\label{prop:pcml}
For finite alphabets $\mathcal{X}$, $\mathcal{Y}$ and $\mathcal{Z}$, the pointwise conditional maximal leakage can be expressed as
\begin{equation}
\mathcal{L}(X\!\to \!Y|Z\!=\!z) = \!\log \sum_{y \in \mathcal{Y}} \max_{x: \, P_{X|Z}(x|z)>0} \!P_{Y|XZ}(y|x,z).
\label{eq:cml2}
\end{equation}
The proof is given in Appendix~\ref{ssec:prop_pcml_proof}.
\end{prop}

Pointwise conditional maximal leakage is an adaptation of conditional maximal leakage proposed in~\cite{issa2019operational} and differs slightly from it. The definition in~\cite{issa2019operational} conditions the leakage on the random variable $Z$ itself, which translates into a maximization over the outcomes of $Z$ in~\eqref{eq:cml2}. We, on the other hand, are conditioning the leakage directly on the outcomes of $Z$ since we are interested in characterizing the leakage for all outcomes, not just the one with the highest leakage. Moreover, as we will see later, the pointwise definition allows us to obtain a data-dependent bound on the leakage which can be tighter than the data-independent bound. More discussions on the comparison of the two bounds can be found in Section~\ref{ssec:ind_leakage}. 

\begin{rem}If the Markov chain $Z-X-Y$ holds,~\eqref{eq:cml2} becomes
\begin{equation}
{\mathcal L}(X \!\to \!Y|Z=z) = \log \sum_{y \in \mathcal{Y}} \max_{x: \, P_{X|Z}(x|z)>0} P_{Y|X}(y|x).
\label{eq:cml3}
\end{equation}
\end{rem}
Similarly to~\cite{issa2019operational}, we now state two important properties of the pointwise conditional maximal leakage: a data-processing inequality and a composition lemma. These properties will be used in the next section to analyze the entrywise information leakage of the PATE framework. 

\begin{lem}[Composition]\label{lemma:composition_pcml}
If the Markov chain $Y_1 \!- \!(X,Z) \!-\! Y_2$ holds, then,
\begin{equation}
\mathcal{L} (X \!\to \!(Y_1,Y_2) \mid Z=z) \leq \mathcal{L} (X \!\to\! Y_1 \mid Z=z) + \mathcal{L} (X \!\to \!Y_2 \mid Z=z).
\end{equation}
More generally, for $k \geq 1$ it holds that 
\begin{equation}
\mathcal{L} (X \!\to \!(Y_1,\ldots ,Y_k) \mid Z=z) \leq \mathcal{L} (X \!\to\! Y_1 \mid Z=z) + \ldots + \mathcal{L} (X \!\to \!Y_k \mid Z=z).
\end{equation}
\end{lem}

Lemma~\ref{lemma:composition_pcml} states that the information leaked to multiple observations is upper bounded by the sum of the information leaked through each of the observations.

\begin{lem}[Data-processing inequality]
\label{lemma:data_processing_pcml}
If the Markov chain $(X,Z) - Y_1 -Y_2$ holds, then,
\begin{equation}
\mathcal{L} (X \! \to \! Y_2 \mid Z=z) \leq \min \{\mathcal{L} (X \! \to \! Y_1 \mid Z=z), \mathcal{L} (Y_1 \! \to \! Y_2 \mid Z=z) \}.
\end{equation}
\end{lem}
Lemma~\ref{lemma:data_processing_pcml} states that all processing of the output can only decrease the information leakage. Further, it allows us to upper bound the end-to-end leakage of a complex mechanism in terms of the leakages of its smaller intermediate mechanisms. The proofs of Lemma~\ref{lemma:composition_pcml} and Lemma~\ref{lemma:data_processing_pcml} are given in Appendix~\ref{ssec:composition_proof} and~\ref{ssec:data_processing_proof}, respectively.  

\section{Information leakage analysis of PATE}
\label{sec:leakage}
In this section, we will use the pointwise conditional maximal leakage to measure the information leaking about individual data entries in the PATE framework. We will begin by describing our system model in Section~\ref{ssec:system_model}. Then, in Section~\ref{ssec:ind_leakage} we will first prove that increased consensus among teachers in answering queries induces a lower privacy cost (i.e., the privacy-accuracy synergy), and then, state bounds on the entrywise leakage when noise with Laplace distribution is used in the aggregation.
\subsection{System Model}
\label{ssec:system_model}
\begin{figure*}
    \centering
    \includegraphics[scale=0.7]{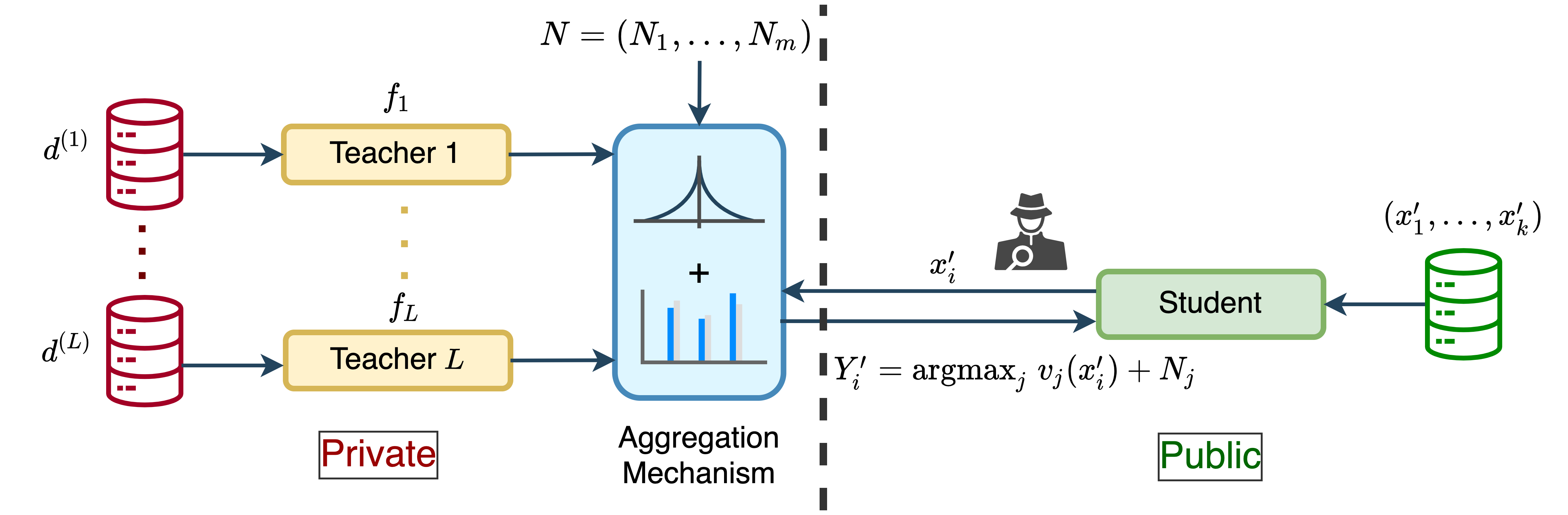}
    \caption{PATE system model~\cite{papernot2016semi}: each partition of the sensitive training data is used to train a teacher. A student model is then trained using a public data-set labelled by the noise-perturbed predictions of the teachers. An adversary who knows all the data-entries except for $D^*$ is trying to guess $D^*$ by observing teachers' responses to queries made by the student.}
    \label{fig:PATE}
\end{figure*}
Suppose $d = ((x_1, y_1), \ldots, (x_n, y_n)) \in \mathcal{X}^n \times \mathcal{Y}^n$ represents the training data where $\mathcal{X}$ is the arbitrary but finite domain set and $\mathcal{Y} = \llbracket 1,m \rrbracket$ is the label set. The pairs $(x_i, y_i)$ are sampled independently according to some distribution $\mathcal{P}$ over $\mathcal{X} \times \mathcal{Y}$, i.e., $D \sim \mathcal{P}^n$. We use the training data $d$ to train $L$ teachers for a classification task with $m \geq 2$ classes in the PATE framework. Let $(d^{(1)}, \ldots, d^{(L)})$ represent a disjoint partitioning of the training set such that $d^{(i)} \neq \emptyset$ for all $i \in \llbracket 1,L \rrbracket$, $\bigcup_{i=1}^L d^{(i)} = d$ and $d^{(i)} \cap d^{(j)} = \emptyset$ for all $i \neq j$. Each partition $d^{(i)}$ is used to train a teacher model $f_i \colon \mathcal{X} \to \llbracket 1,m\rrbracket$. This results in a total of $L$ teacher models, classifying queries independently of each other.

The student model is trained using a public and unlabelled dataset, which will be labelled by the teachers ensemble in a privacy-preserving manner. Let $(x_1', \ldots, x_k') \in \mathcal{X}^k$ be the independently sampled unlabelled dataset and suppose that the student queries the ensemble about the label of $x_i'$. Each teacher separately predicts a label for $x_i'$, referred to as a \textit{vote}. Let $v(x_i') = (v_1(x_i'), \ldots, v_m(x_i'))$ be the histogram of teachers' votes, where $v_j(x_i') = \abs{\{l : l \in \llbracket 1,L\rrbracket, f_l(x_i') = j\}}$ corresponds to the number of teachers who classified $x_i'$ as belonging to class $j$. Note that $\sum_{j=1}^m v_j(x_i') = L$. 

The aggregation mechanism in PATE is essentially the \textit{Report-Noisy-Max mechanism} \cite{dwork2014algorithmic} which operates by adding i.i.d. noise samples to the bins of the votes' histogram, and returning the class label with the highest (noisy) value. Let $\mathrm{Lap}(b)$ denote the Laplace distribution with location $0$ and scale $b$. Suppose $N = (N_1, \ldots, N_m)$ is a sequence of i.i.d. Laplace random variables, where $N_j \sim \mathrm{Lap}({\frac{1}{\gamma}})$ for $j \in \llbracket 1,m \rrbracket$ represents the noise added to the $j$th bin. Note that $\gamma$ determines the dispersion of the noise, and thus, affects the privacy guarantees of the system. Roughly speaking, smaller values of $\gamma$ correspond to larger noise, and in turn, stronger privacy guarantees. Finally, let $Y_i' = \argmax_{j} v_j(x_i') + N_j$ be the random variable denoting the predicted label for $x_i'$ returned by the aggregation mechanism. Labelling the entire dataset $(x_1', \ldots, x_k')$ produces $k$ such predictions, each of which entailing a privacy cost. The system model is depicted in Figure~\ref{fig:PATE}.

\subsection{Measuring the Entrywise Information Leakage}
\label{ssec:ind_leakage}
In this section, we will lay out the details of how we quantify membership privacy through measuring the information leaking about individual data entries in the training set using the notion of pointwise conditional maximal leakage. In order to evaluate the entrywise leakage, let us consider the following scenario: assume an adversary knows the values of all the entries in the teachers' training set (i.e., the private training set) except for a single entry denoted by $D^* = (X^*, Y^*)$. The adversary tries to guess the value of $D^*$ (or any arbitrary discrete function of it) by observing the queries made by the student and their corresponding labels returned by the aggregation mechanism. Clearly, in this setup, observations leak information only about the unknown entry $D^*$ since the adversary already knows all the other entries. 

Now, suppose $(1)$ the adversary has perfect knowledge of the algorithms used to train each teacher, and that $(2)$ the training is done deterministically. That is, we will assume that all classification algorithms and the resulting teacher models (i.e., predictors) are deterministic. Note that the first assumption allows us to remain very conservative about the capabilities of the adversary in order to derive privacy guarantees that remain valid even against highly knowledgeable adversaries. Furthermore, we are using the second assumption to consider a scenario in which the training leaks a lot of information about $D^*$, and the overall privacy guarantees stem only from the aggregation mechanism. As such, our privacy analysis remains valid for all PATE structures regardless of how the teachers are trained, or what classification algorithms are used.

It follows naturally from the previous assumptions that, in principle, the adversary knows all the votes except for the vote of the teacher whose training partition includes $D^*$. Note that we are considering a general setup in which any single data entry can arbitrarily affect the vote of its teacher, resulting in observations which are highly informative for inferring the data entry of interest (as an extreme example, consider a teacher whose vote depends only on $D^*$). In other words, if the adversary can already predict the last vote there is no information left to be leaked. 

Based on the scenario described, let $D^- = D \setminus D^*$ be the random vector representing the portion of the training set known to the adversary, and let $V^-(x_i') = (V_1^-(x_i'), \ldots, V_m^-(x_i'))$ be the random variable representing the histogram of the known votes for input $x_i'$. Note that $\sum_{j=1}^m V_j^-(x_i') = L-1$ for all $x_i' \in \mathcal{X}$. For simplicity, let $Y' = (Y_1', \ldots, Y_k')$ denote the sequence of random variables representing the predicted labels for the queries $(x_1', \ldots, x_k')$. We are interested in quantifying the information leaking about $D^*$ to $Y'$ given that the adversary knows $d^-$ (i.e., the outcome of $D^-$). We have 
\begin{align}
\mathcal{L}(D^* \! \to \! Y' \mid D^- = d^-) &= \! \log \! \sum_{y' \in \mathcal{Y}^k} \max_{\substack{d^* \in \mathcal{X}\times \mathcal{Y} \,:\\ \mathbb{P}(d^* \! \mid d^-)>0}}  \mathbb{P}(y' \mid d^*, d^-) \nonumber \\
& = \! \log \! \sum_{y' \in \mathcal{Y}^k} \max_{\substack{d \in \mathcal{X}^n \times \mathcal{Y}^n \,:\\ \mathbb{P}(d \mid d^-)>0}}  \mathbb{P}(y' \! \mid \! d) \\
&\labelrel={eq:ind_leakage_p1} \mathcal{L}(D \! \to \! Y' \mid D^-=d^-), \nonumber
\label{eq:ind_leakage}
\end{align}
where~\eqref{eq:ind_leakage_p1} follows from~\eqref{eq:cml3} since the Markov chain ${D^- \!-\! D \!-\! Y'}$ holds. Using Lemma~\ref{lemma:composition_pcml} we can upper bound the information leaked through multiple queries by writing 
\begin{equation}
\mathcal{L}(D \!\to \! Y' \! \mid \! D^-=d^-) \leq \sum_{i=1}^k \mathcal{L}(D \! \to \! Y_i' \! \mid \! D^-=d^-),
\end{equation} 
that is, the information leaked to the output of multiple queries is upper bounded by the sum of the information leaked through individual queries. Further, using Lemma~\ref{lemma:data_processing_pcml} we can upper bound the information leaked to the output of a single query as 
\begin{equation}
\mathcal{L}(D \! \to \! Y_i' \mid D^-=d^-) \leq \min \{\mathcal{L}(D \! \to \! V(x_i') \mid D^-=d^-), \mathcal{L}(V(x_i') \! \to \! Y_i' \mid D^-=d^-)\},
\end{equation}
i.e., the information leaked to the output of a single query is upper bounded by the smallest of the information leaked through the training and the information leaked through the aggregation mechanism. 

As we do not want to make any assumptions about how privately the teachers are trained, we now turn to evaluating the information leaked through the aggregation mechanism. Let ${\delta_j = (0, \ldots, 0, 1, 0, \ldots, 0)}$ be a sequence with all elements equal to $0$, except for the $j$th element which equals $1$. We will use $\delta_j$ to represent a single vote for class $j$. Then, we have
\begin{align}
\label{eq:aggr_leakage}
\mathcal{L}(V(x_i') \! \to \! Y_i' \! \mid \! D^-=d^-) &= \mathcal{L}(V(x_i') \! \to \! Y_i' \! \mid \! V^-(x_i')=v^-) \nonumber \\
&\labelrel={eq:aggr_leakage_p1} \! \log \! \sum_{j=1}^m  \max_{\substack{v = v^- + \delta_{j'} :\\ j' \in \llbracket 1,m\rrbracket}} \! \! \mathbb{P}(Y_i'\!=\!j \! \mid \! V(x_i') \!= \!v)  \\ 
&\labelrel={eq:aggr_leakage_p2} \log \sum_{j=1}^m \mathbb{P}(Y_i'=j \! \mid \! V(x_i') = v^- + \delta_j),\nonumber
\end{align}
where~\eqref{eq:aggr_leakage_p1} follows from~\eqref{eq:cml3}, and~\eqref{eq:aggr_leakage_p2} follows from the fact that the probability of outputting class $j$ is maximized when the last vote (i.e., the vote of the teacher whose training partition includes $D^*$) is placed for class $j$. 

\subsubsection{The privacy-accuracy synergy}
Now, we will evaluate the leakage of the aggregation mechanism as described by~\eqref{eq:aggr_leakage} using ideas from majorization theory~\cite{marshall1979inequalities, jorswieck2007majorization} and assuming that the noise used in the mechanism has a log-concave probability density~\cite{bagnoli2005log, an1997log}. Specifically, we will find the $v^-$ maximizing or minimizing~\eqref{eq:aggr_leakage} for any noise with log-concave probability density.
\begin{defi}[Majorization]\label{def_majorization}
Consider $p, q \in \mathbb{R}^n$ with non-increasingly ordered elements, i.e., $p_1 \geq p_2 \geq \ldots \geq p_n$ and $q_1 \geq q_2 \geq \ldots \geq q_n$. We say that $p$ majorizes $q$, and write $p \succ q$ if 
\begin{equation}
\sum_{i=1}^m \!p_i \geq \sum_{i=1}^m \!q_i, \; \text{for} \; m\!=\!1, \! \ldots, n\!-\!1 \; \text{and} \;  \sum_{i=1}^n \! p_i \!=\! \sum_{i=1}^n \!q_i.
\end{equation}
\end{defi}
Note that majorization only describes a partial ordering. For example, $(4,4,1)$ and $(5,2,2)$ cannot be compared in terms of majorization. On the other hand, if we define $\mathcal{Q} = \{(q_1, q_2, q_3) \in \mathbb{R}_+^3 : \sum_{i=1}^3 q_i = 9\}$, then $(3,3,3)$ is majorized by all $q \in \mathcal{Q}$ while $(9,0,0)$, $(0,9,0)$ and $(0,0,9)$ majorize all $q \in \mathcal{Q}$. 

\begin{defi}[Schur-concave function]\label{def_schur_concave}
Consider a real-valued function $\Phi$ defined on $\mathcal{I}^n \subset \mathbb{R}^n$. $\Phi$ is said to be Schur-concave on $\mathcal{I}^n$ if $p \succ q $ on $\mathcal{I}^n$ implies $\Phi(p) \leq \Phi(q)$. 
\end{defi}  

\begin{defi}[Log-concave function]\label{def_log_concave}
A non-negative function $f : \mathbb{R}^n \to \mathbb{R}_+$ is said to be log-concave if it can be written as $f(x) = \exp \phi(x)$ for some concave function ${\phi : \mathbb{R}^n \to [-\infty, \infty)}$.
\end{defi}

Note that many commonly used probability density functions (and their corresponding CDFs) are log-concave, such as the Laplace and the Gaussian distributions~\cite{bagnoli2005log}.  

\begin{theo}\label{theo:max_individual_leakage}
Consider the aggregation mechanism in PATE (i.e., the Report-Noisy-Max mechanism) where the noise has a log-concave probability density. Then, ${\mathcal{L}(V(x_i') \! \to \! Y_i' \! \mid \! V^-(x_i') = v^-)}$ is Schur-concave in $v^-$. Thus, assuming that $L-1$ is divisible by $m$, ${\mathcal{L}(V(x_i') \! \to \! Y_i' \! \mid \!V^-(x_i') = v^-)}$ is maximized when 
\begin{equation}\label{eq:z_unif}
v^-= v_{max}^- = \left(\frac{L-1}{m}, \ldots, \frac{L-1}{m}\right),
\end{equation}
and is minimized when 
\begin{equation}
v^- \!=\! v_{min}^- \!=\! (0, \ldots, 0, L-1, 0, \ldots, 0) = (L-1)\, \delta_j,
\end{equation}
for some $j \in \llbracket 1, m\rrbracket$.
\end{theo}
The proof of the theorem is given in Appendix~\ref{sec:proof_theorem1}. 

\begin{rem}\label{rem:schur-concavity}
The Schur-concavity of the entrywise information leakage of the aggregation mechanism $\mathcal{L}(V(x_i') \! \to \! Y_i' \! \mid \! V^-(x_i') = v^-\!)$ implies that stronger consensus among teachers lowers the amount of information leaked about any individual data entry.
\end{rem}
 
The preceding remark points to one of the main advantages of the PATE framework: increased accuracy of the teacher models results in stronger consensus in predicting the label of a given query, which, in turn, results in stronger privacy guarantees. Note that~\cite{papernot2016semi, papernot2018scalable} intuitively come to the same conclusions regarding the synergy between privacy and accuracy for the case of Laplace and Gaussian noise distributions, whereas here we have analytically proved this property and generalized it to the class of log-concave probability densities.

\subsubsection{Data-independent bound}
Now, we will apply Theorem~\ref{theo:max_individual_leakage} to~\eqref{eq:aggr_leakage} to get a bound on the leakage  of the aggregation mechanism with Laplace noise.  
\begin{prop}\label{prop:data_independent}
Consider the PATE framework where noise with Laplace distribution is used in the aggregation mechanism. For all $v^-$, the information leaked to the output of a single query is upper bounded by
\begin{equation}
\mathcal{L}(V(x_i')  \to  Y_i'  \mid  V^-(x_i')=v^-) \leq \log (B_1), 
\end{equation}
where 
\begin{equation}
B_1 \coloneqq (1-m)\, 2^{-m} e^{-\gamma} + e^{\gamma} \left(1-{(1-\frac{1}{2}e^{-\gamma})}^m\right) + \frac{m}{2} {(1-\frac{1}{2}e^{-\gamma})}^{m-1} - \frac{m(m-1)}{4} e^{-\gamma}  H(m-2).
\end{equation}
Also, $H(0) \coloneqq \gamma$ and 
\begin{equation}\label{eq:h(m)}
H(m) \coloneqq \gamma + \sum_{k=1}^m \frac{2^{-k} - {(1-\frac{1}{2}e^{-\gamma })}^{k}}{k} \quad \text{for} \;\; m \geq 1, 
\end{equation}
The bound is attained at $v^- = v_{max}^-$ defined in~\eqref{eq:z_unif}.
\end{prop}

The proof of this result is given in Appendix~\ref{ssec:proof_prop2}. Proposition~\ref{prop:data_independent} describes a data-independent bound that holds uniformly for all $v^-$ (and consequently all $d^-$) but depends on $m$, the number of classes. It can be verified through simple calculations that the bound is non-decreasing in $m$. Therefore, by letting $m$ tend to infinity, we get the following simpler bound which holds for all $d^-$ and all $m \geq 2$. 

\begin{theo}\label{theo:ind_leakage}
Consider the setting of Proposition~\ref{prop:data_independent}. For all $d^-$ and all $m \geq 2$, the information leaked about $D^*$ as a result of labelling a single query is upper bounded by
\begin{align}
    \mathcal L (D^* \to Y_i' \mid D^- = d^-) &= {\mathcal L}(D \to Y_i' \mid D^-=d^-) \\\nonumber
    & \leq \mathcal{L}(V(x_i')  \to  Y_i'  \mid  V^-(x_i')=v^-)\\\nonumber
    &\leq \gamma. 
\end{align}
The proof of this result is given in Appendix~\ref{ssec:proof_theo2}. 
\end{theo}

Note that the bounds stated in Proposition~\ref{prop:data_independent} and Theorem~\ref{theo:ind_leakage} give a more accurate characterization of the leakage as consensus among teachers decreases. This is demonstrated in the following example where we calculate the leakage in~\eqref{eq:aggr_leakage} directly using the conditional probabilities, and compare it with the bounds.

\begin{exe}
Suppose the PATE framework has been implemented with $L=11$ teachers to classify queries into $m=4$ classes. Further, suppose that for a given query $x_i'$, the histogram of teachers' votes is (some permutation of) $v = (5,3,2,1)$, and that Laplace noise with $\gamma = 0.1$ is used in the aggregation mechanism. Depending on which partition of the training set includes $D^*$, the adversary has obtained one of the following values: $v^- \in \{(4, 3, 2, 1), (5, 2, 2, 1), (5, 3, 1, 1), (5, 3, 2, 0)\}$. We can now directly use~\eqref{eq:aggr_leakage} to calculate the leakage of the aggregation mechanism using the probability density function of the Laplace distribution for each $v^-$. For simplicity of notation we define $\mathcal{L}(v^-) \coloneqq \mathcal{L}(V(x_i') \! \to \! Y_i' \mid V^-(x_i')=v^-)$. Then, we have one of the following four cases:
\begin{itemize}
\item $v^- \! = \! (4, 3, 2, 1) \! \implies \! \mathcal{L}(v^-) = 8.50 \times 10^{-2}$.
\item $v^- \! = \! (5, 2, 2, 1) \! \implies \! \mathcal{L}(v^-) = 8.40 \times 10^{-2}$.
\item $v^- \! = \! (5, 3, 1, 1) \! \implies \! \mathcal{L}(v^-) = 8.37 \times 10^{-2}$.
\item $v^- \! = \! (5, 3, 2, 0) \! \implies \! \mathcal{L}(v^-) = 8.35 \times 10^{-2}$.
\end{itemize}
Therefore, $\mathcal{L}(v^-) \leq 8.50 \times 10^{-2}$ while Proposition~\ref{prop:data_independent} predicts $\mathcal{L}(v^-) \leq \log(B_1) = 8.61 \times 10^{-2}$ and Theorem~\ref{theo:ind_leakage} predicts $\mathcal{L}(v^-) \leq 0.1$. Note that due to the Schur-concavity of $\mathcal{L}(v^-)$ it was already expected that information leakage would be largest for $(4, 3, 2, 1)$, and it would have sufficed to just consider this case. Now, suppose $v = (3,3,3,2)$. Calculating the leakage using the corresponding conditional probabilities gives $\mathcal{L}(v^-) \leq 8.58 \times 10^{-2}$, which is closer to the value predicted by Proposition~\ref{prop:data_independent} and Theorem~\ref{theo:ind_leakage}.
\end{exe} 

Our final data-independent bound describes the information leaked through multiple queries.

\begin{cor}
Consider the setting of Theorem~\ref{theo:ind_leakage}. The information leaked about $D^*$ as the result of training a student model on $k$ samples is upper bounded by  
\begin{equation}
\mathcal L (D^* \! \to \! Y' \mid D^- = d^-) \leq k \gamma. 
\end{equation}
\end{cor}
This result is a direct consequence of Theorem~\ref{theo:ind_leakage} and Lemma~\ref{lemma:composition_pcml}, and characterizes the overall information leaked about a single data entry as a result of training a student classifier using $k$ queries to the teachers. 

\subsubsection{Data-dependent bound}
In the previous section, we presented bounds on the leakage that hold uniformly regardless of the data used in the training. Here, we present a bound that depends on the training data through $v^-$.  

\begin{prop}\label{prop:data_dependent}
Consider the PATE framework where noise with Laplace distribution is used in the aggregation mechanism. Suppose $v^-$ is sorted in non-increasing order and that the first $r$ coordinates have equal votes, that is, $v^-_1 = \ldots = v^-_r > v^-_{r+1} \geq \ldots \geq v^-_m$ for some $1 \leq r \leq m$. Then, we have 
\begin{equation}
    \mathcal{L}(V(x_i')  \to  Y_i'  \mid  V^-(x_i')=v^-) \leq \; \log (B_2),
\end{equation}
where
\begin{equation}
    B_2 \coloneqq r \left(1 - \frac{2 + \gamma (v^-_1 + 1 - v^-_2)}{4 \exp\left(\gamma(v^-_1 + 1 - v^-_2)\right)} \right) + \sum_{j=r+1}^m \frac{2 + \gamma (v^-_1 - 1 - v^-_j)}{4 \exp \left( \gamma (v^-_1 - 1 - v^-_j)\right)}. \hspace{0.5cm}
\end{equation}
The proof of this result is given in Appendix~\ref{ssec:proof_prop_data_dependent}.
\end{prop}

In practice, in order to calculate the information leaked through a query response, one has to take the minimum of the data-dependent bound in Proposition~\ref{prop:data_dependent} and the data-independent bound in Proposition~\ref{prop:data_independent}. Roughly speaking, the data-dependent bound is tighter than the data-independent bound when the teachers have strong agreement over the label of a query. This is illustrated in the numerical example below. 

\begin{exe}
Suppose the PATE framework has been implemented for a classification task with $m=4$ classes and that Laplace noise with $\gamma = 0.1$ is used in the aggregation mechanism. First, consider the case where $L=11$ and $v^- = (4,3,2,1)$. Then, $\log (B_1) = 8.61 \times 10^{-2}$ while $\log (B_2) = 6.81 \times 10^{-1}$, so the data-independent bound is much tighter. Now, suppose $L=101$ and $v^- = (90,5,5,0)$. Then, $\log (B_2) = 1.05 \times 10^{-3}$, while the data-independent remains as before. Therefore, the data-dependent bound is tighter when there is a strong consensus among teachers. 
\end{exe}

\section{Conclusions}
\label{sec:conclusions}
In this paper, we have proposed an approach based on information leakage for quantifying membership privacy. Particularly, we showed that the pointwise conditional maximal leakage, a conditional form of maximal leakage, can be used to measure the information leaking about individual data entries in a dataset. We applied our privacy analysis to PATE and derived novel privacy guarantees for this privacy-preserving classification framework in the form of upper bounds on its entrywise information leakage when the injected noise has Laplace distribution. We also showed that the privacy-accuracy synergy of PATE can be explained by studying the entrywise information leakage of the framework while it was only intuitively justified through the lens of differential privacy.  

As our work has taken a step towards gaining a deeper understanding of some underlying privacy principles in the PATE framework, our results can be used in the design of machine learning algorithms that preserve both privacy and utility. For example, we can consider a situation in which we have a fixed privacy budget per query. Then, using the data-dependent bound of Proposition~\ref{prop:data_dependent}, one can adjust the noise parameter $\gamma$ in order to achieve the budget for each query. We except that this will improve the utility of the system since, for example, less noise will be required when there is a strong consensus over the label of a query. Another potential application is in privacy thresholding schemes where queries which are expensive in terms of privacy will not be answered at all. Once again this method will improve both the privacy and the utility of the system since the expensive queries are precisely those which were not labelled with certainty by the teachers.


%

\appendices
\section{Proofs of the results in section~\ref{sec:pcml_properties}}
\subsection{Proof of Proposition~\ref{prop:pcml}}\label{ssec:prop_pcml_proof}
This result follows readily from~\cite[Theorem 1]{issa2019operational} by considering $\mathcal{L}(X' \! \to \! Y)$ such that $P_{X'} = P_{X \mid Z=z}$. Nevertheless, we provide an alternative proof.

\textbf{Upper bound:} \space First, we prove the upper bound on ${\mathcal{L}(X \! \to \! Y \mid Z=z)}$. Consider any discrete $U$ satisfying ${U-(X,Z)-Y}$ and define  
\begin{equation}
\mathcal{L}_U(X \! \to \! Y \mid Z=z) \coloneqq \log{ \frac{\mathbb{P} \left(U= \hat U(Y,Z=z)\right)}{\mathbb{P} \left(U= \tilde U(Z=z)\right)}},
\label{eq:leakage_fixed_u}
\end{equation}
where $\hat{U}$ and $\tilde{U}$ are MAP estimators of $U$. Then, ${\mathcal{L}(X \! \to \! Y \! \mid \! Z=z) = \sup_{U:U-(X,Z)-Y} \mathcal{L}_U(X \! \to \! Y \! \mid \! Z=z)}$.

For each $z \in \mathcal{Z}$, define $\mathcal{U}_z \coloneqq \{u : P_{U \mid Z} (u \mid z) >0 \}$. The two probabilities in $\mathcal{L}_U(X \! \to \! Y \! \mid \! Z=z)$ are 
\begin{equation}
\mathbb{P} \left(U= \tilde U(Z=z)\right) = \max_{u \in \mathcal{U}_z} P_{U \mid Z}(u \mid z),
\end{equation}
and 
\begin{align}
\begin{split}
\mathbb{P} \left(U= \hat U(Y,Z=z)\right) &= \sum_{y \in \mathcal{Y}} \max_{u \in \mathcal{U}_z} P_{UY \mid Z} (u,y \mid z) \\
&= \! \sum_{y \in \mathcal{Y}} \max_{u \in \mathcal{U}_z} \! \! \sum_{x:P_{X \mid Z} (x \mid z) > 0}\hspace{-7mm} P_{U \mid XZ} (u \! \mid \! x,z) P_{Y \mid XZ} (y \! \mid \! x,z) P_{X \mid Z} (x \! \mid \! z) \\
& \leq \! \sum_{y \in \mathcal{Y}} \! \! \left( \! \max_{x':P_{X \mid Z} (x' \mid z) > 0} \! \! P_{Y \mid XZ} (y \! \mid \! x,z) \! \right) \! \max_{u \in \mathcal{U}_z} \hspace{-4mm} \sum_{ \, x:P_{X \mid Z} (x \mid z) > 0} \hspace{-7mm} P_{UX \mid Z} (u,x \!\! \mid \!\! z) \\
&= \max_{u \in \mathcal{U}_z} P_{U  \mid  Z} (u \! \mid \! z) \sum_{y \in \mathcal{Y}} \! \left(\max_{x':P_{X \mid Z} (x' \mid z) > 0} \! P_{Y \mid XZ} (y \! \mid \! x,z) \! \right). \\
\end{split}
\end{align}
Thus,
\begin{align}
\mathcal{L}_U(X \! \to \! Y \! \mid \! Z=z) \leq \log \! \sum_{y \in \mathcal{Y}} \max_{x: \, P_{X|Z}(x|z)>0} \! P_{Y|XZ}(y \! \mid \! x,z)
\end{align}
for all $U$ such that $U -(X,Z) - Y$ holds. Then,
\begin{equation}
\mathcal{L}(X \to Y \mid Z=z) \leq \log \sum_{y \in \mathcal{Y}} \max_{x: \, P_{X|Z}(x|z)>0} P_{Y|XZ}(y|x,z).
\label{eq:upper_bound_pcml}
\end{equation}

\textbf{Lower bound}: \space To prove the lower bound on ${\mathcal{L}(X \! \to \! Y \! \mid \! Z=z)}$, we will consider a discrete $U$ for which ${\mathcal{L}_U(X \! \to \! Y \! \mid \! Z=z)}$ attains the bound. We fix a $U'$ such that $U' -(X,Z) - Y$ holds and $H(X \! \mid \! U') =0$, that is, the value of $X$ is completely determined by the value of $U'$. Further, we assume that $U'\! \mid \! Z\! =\! z$ is uniformly distributed, i.e., ${P_{U'\mid Z}(u \! \mid \! z) = \frac{1}{\lvert \mathcal{U}_z \rvert}}$ for all $z \in \mathcal{Z}$ and $u \in \mathcal{U}_z$. Then, 
\begin{align}
\begin{split}
   \mathcal{L}_{U'}(X \to Y \mid Z=z) &= \log \sum_{y \in \mathcal{Y}} \frac{\max_{u \in \mathcal{U}_z} P_{U '\mid Z}(u \mid z) P_{Y \mid U'Z}(y \mid u, z)}{\max_{u \in \mathcal{U}_z} P_{U' \mid Z}(u \mid z)} \\[3mm] 
    &= \log \sum_{y \in \mathcal{Y}} \frac{\max_{u \in \mathcal{U}_z} P_{U' \mid Z}(u \mid z) P_{Y \mid XZ} (y \mid x, z)}{\max_{u \in \mathcal{U}_z} P_{U' \mid Z}(u \mid z)} \\[3mm] 
    &= \log \sum_{y \in \mathcal{Y}} \frac{\frac{1}{\lvert \mathcal{U}_z \rvert} \max_{x: \, P_{X|Z}(x|z)>0} P_{Y \mid XZ} (y \mid x, z)}{\frac{1}{\lvert \mathcal{U}_z \rvert}} \\[3mm] 
    &= \log \sum_{y \in \mathcal{Y}} \max_{x: \, P_{X|Z}(x|z)>0} P_{Y|XZ}(y|x,z). 
\end{split}
\label{eq:proof_lower_bound_pcml}
\end{align}
Therefore, 
\begin{align}
\begin{split}
\mathcal{L}(X \to Y \mid Z=z) &= \! \sup_{U:\, U-(X,Z)-Y} \mathcal{L}_U(X \! \to \! Y \! \mid \! Z=z) \\ 
&\geq \mathcal{L}_{U'}(X \! \to \! Y \! \mid \! Z=z) \\ 
&= \log \sum_{y \in \mathcal{Y}} \max_{x: \, P_{X|Z}(x|z)>0} \! \! P_{Y|XZ}(y|x,z).
\end{split}
\label{eq:lower_bound_pcml}
\end{align}
Hence, from~\eqref{eq:upper_bound_pcml} and~\eqref{eq:lower_bound_pcml} it follows that 
\begin{equation}
\mathcal{L}(X \! \to \! Y \! \mid \! Z=z) = \! \log \sum_{y \in \mathcal{Y}} \! \max_{x: \, P_{X|Z}(x|z)>0} P_{Y|XZ}(y|x,z).
\end{equation}

\subsection{Proof of Lemma~\ref{lemma:composition_pcml}}\label{ssec:composition_proof}
Consider the Markov chain $Y_1 -(X,Z)- Y_2$. Then, 
\begin{align}
\begin{split}
    \mathcal{L}(X \! \to \! (Y_1, Y_2) \! \mid \! Z=z) - \mathcal{L}(X \! \to \! Y_1 \! \mid \! Z\! =\! z) &= \! \log \! \frac{\sum_{y_1, y_2} \max_{x:P_{X \mid Z}(x \mid z)>0} P_{Y_1 Y_2 \mid X Z}(y_1, y_2 \! \mid \! x,z)}{\sum_{y_1} \max_{x:P_{X \mid Z}(x \mid z)>0} P_{Y_1 \mid X Z}(y_1 \! \mid \! x,z)} \\[3mm]
    &= \! \log \! \sum_{y_2} \! \frac{\sum_{y_1} \! \! \max_{x} \! P_{Y_1 \mid X Z}(y_1 \! \mid \! x,z)  P_{Y_2 \mid X Z}(y_2 \! \mid \! x,z)}{\sum_{y_1} \max_{x:P_{X \mid Z}(x \mid z)>0} P_{Y_1 \mid X Z}(y_1 \mid x,z)} \\[3mm]
    &\leq \! \log \! \sum_{y_2} \! \frac{\sum_{y_1} \! \! \max_{x} P_{Y_1 \mid X Z}(y_1 \! \mid \! x,z) \! \left(\max_{x'} P_{Y_2 \mid X Z}(y_2 \! \mid \! x',z)\right)}{\sum_{y_1} \max_{x:P_{X \mid Z}(x \! \mid \! z)>0} P_{Y_1 \mid X Z}(y_1 \mid x,z)} \\[3mm]
    &= \! \log \! \sum_{y_2} \max_{x':P_{X \mid Z}(x' \mid  z)>0} P_{Y_2 \mid X Z}(y_2 \! \mid  \! x',z) \\[3mm]
    &= \mathcal{L}(X \to Y_2 \mid Z=z).  
\end{split}
\end{align}
Therefore, 
\begin{equation}
\mathcal{L} (X \to (Y_1,Y_2) \mid Z=z) \\ 
\leq \mathcal{L} (X \to Y_1 \mid Z=z) + \mathcal{L} (X \to Y_2 \mid Z=z).
\end{equation}

\subsection{Proof of Lemma~\ref{lemma:data_processing_pcml}}\label{ssec:data_processing_proof}
Our proof follows the same reasoning as the proof of~\cite[Lemma 1]{issa2019operational}. For all discrete $U$ satisfying $U-(X,Z)-Y_1 - Y_2$ it holds that 
\begin{equation}
\mathcal{L}_U(X \! \to \! Y_2 \! \mid \! Z=z) \leq \mathcal{L}_U(X \! \to \! Y_1 \! \mid \! Z=z),
\end{equation}
where $\mathcal{L}_U$ is defined in~\eqref{eq:leakage_fixed_u}. Therefore, 
\begin{align}
\begin{split}
    \mathcal{L}(X \! \to \! Y_2 \! \mid \! Z=z) &= \sup_{U:\, U-(X,Z)-Y_1-Y_2} \! \! \mathcal{L}_U(X \! \to \! Y_2 \! \mid \! Z=z) \\
    &\leq \sup_{U:\, U-(X,Z)-Y_1} \! \mathcal{L}_U(X \! \to \! Y_1 \! \mid \! Z=z) \\
    &= \mathcal{L}(X \! \to \! Y_1 \! \mid \! Z=z).
\end{split}
\end{align}
Similarly, 
\begin{align}
\begin{split}
    \mathcal{L}(X \! \to \! Y_2 \! \mid \! Z=z) &=  \sup_{U:\, U-(X,Z)-Y_1-Y_2} \! \! \mathcal{L}_U(X \! \to \! Y_2 \! \mid \! Z=z) \\
    &\leq \sup_{U:\, U-Z-Y_1-Y_2} \! \! \mathcal{L}_U(Y_1 \! \to \! Y_2 \! \mid \! Z=z) \\
    &= \mathcal{L}(Y_1 \! \to \!  Y_2 \! \mid \! Z=z).
\end{split}
\end{align}
Thus, 
\begin{equation}
\mathcal{L} (X \! \to \! Y_2 \! \mid \! Z=z) \leq \min \{\mathcal{L} (X \! \to \! Y_1 \! \mid  \! Z=z), \\
 \mathcal{L} (Y_1 \! \to  \! Y_2 \! \mid \!  Z=z) \}.
\end{equation}
\section{Proof of Theorem~\ref{theo:max_individual_leakage}}\label{sec:proof_theorem1}
Before stating the proof, let us recall some concepts/results from majorization theory. 
\begin{defi}[Symmetric function] 
Let $x \! = \! (x_1, \ldots, x_n) \! \in \! \mathcal{I}^n \! \subset \! \mathbb{R}^n$ and consider a real-valued function $\Phi : \mathcal{I}^n \to \mathbb{R}$. The function $\Phi(x)$ is said to be symmetric if $x$ can be arbitrarily permuted without changing the value of $\Phi(x)$. 
\end{defi}

\begin{lem}[Schur's condition]
Let $x \! = \! (x_1, \ldots, x_n) \! \! \in \! \mathcal{I}^n \! \subset \! \mathbb{R}^n$ and consider a continuously differentiable function ${\Phi : \mathcal{I}^n \to \mathbb{R}}$. $\Phi(x)$ is Schur-concave on $\mathcal{I}^n$ if and only if it is symmetric on $\mathcal{I}^n$ and 
\begin{equation}
(x_i - x_j)\left(\frac{\partial f}{\partial x_i} - \frac{\partial f}{\partial x_j}\right) \leq 0 \quad \text{for all} \quad 1 \leq i,j \leq n.
\label{eq:Schurs_condition}
\end{equation}
Since $\Phi(x)$ must be symmetric, it is sufficient to verify the reduced condition
\begin{equation}
(x_1 - x_2)\left(\frac{\partial f}{\partial x_1} - \frac{\partial f}{\partial x_2}\right) \leq 0.
\label{eq:simple_Schurs_condition}
\end{equation} 
\end{lem}

\begin{prop}[{\cite[Theorem 2.21]{jorswieck2007majorization}}]
Let $x \! = \! (x_1, \ldots, x_n) \! \in \! \mathbb{R}_+^n$ and let $f: \mathbb{R}_+^n \to \mathbb{R}_+$ be a Schur-concave function. Consider the following problems 
\begin{equation}
\max_{x} f(x) \quad \text{such that} \quad \; \sum_{i=1}^n x_i = S, 
\end{equation}
and 
\begin{equation}
\min_{x} f(x) \quad \text{such that} \quad \; \sum_{i=1}^n x_i = S. 
\end{equation}
Then, the global maximum is achieved by 
\begin{equation}
x_{max} = \frac{S}{n}(1, \ldots, 1),
\end{equation}
and the global minimum is achieved by 
\begin{equation}
x_{min} = (0, \ldots, 0, S, 0, \ldots, 0).
\end{equation}
\label{prop:max_min_schur_concave}
\end{prop}

We now prove that the entrywise information leakage of the aggregation mechanism is Schur-concave when the injected noise has a log-concave probability density. In order to simplify the proof, we will assume that the elements of $v^-$ (i.e., the histogram of known votes) can take non-negative real values. The results of the proof, however, will be readily applicable to histograms of non-negative integers. 

Using~\eqref{eq:aggr_leakage} we define 
\begin{equation}
f_j(v^-) \coloneqq \mathbb{P}(Y_i'=j \mid V(x_i') = v^- + \delta_j),
\end{equation}
where $\delta_j = (0, \ldots, 0, 1, 0, \ldots, 0)$  represents a single vote for class $j$. Then, 
\begin{equation}
\mathcal{L}(V(x_i') \to Y_i' \mid V^-(x_i')=v^-) = \log \sum_{j=1}^m f_j(v^-) = \log f(v^-),
\label{eq:leakage_as_sum}
\end{equation}
where $f(v^-) = \sum_{j=1}^m f_j(v^-)$. 
It is clear from~\eqref{eq:leakage_as_sum} that the leakage does not depend on the order of elements in $v^-$, thus ${\mathcal{L}(V(x_i') \! \to \! Y_i' \! \mid \! V^-(x_i')=v^-)}$ is symmetric. Moreover, according to~\cite[3.B.1]{marshall1979inequalities}, the composition of an increasing function and a Schur-concave function remains Schur-concave. Since $\log(\cdot)$ is an increasing function, to prove the Schur-concavity of the entrywise leakage we only need to verify Schur's condition for $f(v^-)$.

Without loss of generality assume that ${v^- = (v^-_1, \ldots, v^-_m)}$ is non-increasingly ordered, i.e., ${v^-_1 \geq \ldots \geq v^-_m}$. Let ${N=(N_1, \ldots, N_m)}$ denote the tuple of noise, where the elements are independent, identically distributed and have a log-concave probability density. We write
\begin{align}
\begin{split}
    f_j(v^-) &= \mathbb{P}(Y_i'=j \! \mid \! V(x_i') = v^- + \delta_j)\\[2mm]
    &= \mathbb P \{ v^-_j \! +\! N_j \! + \! 1> v^-_1\!+\!N_1, \ldots, v^-_j\!+\!N_j\!+\!1> v^-_m\!+\!N_m\} \\[1mm]
    &= \int_{-\infty}^{\infty} \left[ \prod_{\substack{l=1\\ l \neq j}}^m \mathbb P \{N_l <(v^-_j -v^-_l +t +1) \} \right]  g(t) dt \\
    &= \int_{-\infty}^{\infty} \left[ \prod_{\substack{l=1\\ l \neq j}}^m G(v^-_j -v^-_l +t +1) \right] g(t) \, dt,
\end{split}
\end{align}
where $g(t)$ is the probability density function of $N_j$ and $G(t) = \int_{-\infty}^{t} g(t') \, dt'$ is its corresponding cumulative distribution function. According to~\cite[Proposition 1]{an1997log}, if $g$ is log-concave, then $G$ is also log-concave. We now check Schur's condition by writing 
\begin{equation}
\frac{\partial f(v^-)}{\partial v^-_1} - \frac{\partial f(v^-)}{\partial v^-_2} = \sum_{j=1}^m \frac{\partial f_j(v^-)}{\partial v^-_1} - \frac{\partial f_j(v^-)}{\partial v^-_2},
\end{equation}
where we have one of the following three cases: \\
if $j=1$, then, 
\begin{align}
\begin{split}
\frac{\partial f_1(v^-)}{\partial v^-_1} - \frac{\partial f_1(v^-)}{\partial v^-_2} &= \! \int_{-\infty}^{\infty} \! \left[ \sum_{l=2}^m g(v^-_1\!-\!v^-_l\!+\!t\!+\!1) \prod_{\substack{k=2 \\ k \neq l}}^m G(v^-_1\!-\!v^-_k\!+\!t\!+\!1) \! \right] \! g(t) \, dt \\ 
& \hspace{-5mm} - \! \int_{-\infty}^{\infty} \! \left[  - g(v^-_1\!-\!v^-_2\!+\!t\!+\!1) \right] \! \left[ \prod_{l=3}^m G(v^-_1\!-\!v^-_l\!+\!t\!+\!1)\! \right] \! g(t) \, dt,
\end{split}
\end{align}
if $j=2$, then, 
\begin{align}
\begin{split}
\frac{\partial f_2(v^-)}{\partial v^-_1} - \frac{\partial f_2(v^-)}{\partial v^-_2} &= \! \int_{-\infty}^{\infty} \! \left[- g(v^-_2\!-\!v^-_1\!+\!t\!+\!1) \right] \! \left[ \prod_{l=3}^m G(v^-_2-v^-_l\!+\!t\!+\!1) \right] \! g(t) \, dt  \\
& \hspace{-5mm} - \! \int_{-\infty}^{\infty} \! \left[ \sum_{\substack{l=1 \\ l \neq 2}}^m  g(v^-_2\!-\!v^-_l\!+\!t\!+\!1) \! \! \! \prod_{\substack{k=1\\ k\neq 2,l}}^m \! \! G(v^-_2\!-\!v^-_k\!+\!t\!+\!1)\!  \right] \! g(t) \, dt,
\end{split}
\end{align}
and if $j \neq 1,2$, then, 
\begin{align}
\frac{\partial \! f_j(v^-)}{\partial v^-_1} - \frac{\partial \! f_j(v^-)}{\partial v^-_2} &= \int_{-\infty}^{\infty} - \! \bigg[ g(v^-_j\!-\!v^-_1\!+\!t\!+\!1) G(v^-_j\!-\!v^-_2\!+\!t\!+\!1) \\
& \hspace{-5mm} + \! g(v^-_j\!-\!v^-_2\!+\!t\!+\!1) \, G(v^-_j\!-\!v^-_1\!+\!t\!+\!1)\bigg] \cdot \left[\prod_{\substack{l=3\\ l \neq j}}^m G(v^-_j-v^-_l+t+1)\right] \, g(t) \, dt. \nonumber
\end{align}

Then, 
\begin{equation}
\frac{\partial f(v^-)}{\partial v^-_1} - \frac{\partial f(v^-)}{\partial v^-_2}= A_1 - A_2 + \sum_{j=3}^{m} B_{(1, j)} - B_{(2, j)} + B_{(3, j)} - B_{(4, j)},
\end{equation}
where 
\begin{equation}
A_1\!=\!2\!\int_{-\infty}^{\infty}  \! \! g(v^-_1\!-\!v^-_2\!+\!t\!+\!1)\! \left[\prod_{k=3}^m G(v^-_1\!-\!v^-_k\!+\!t\!+\!1)\right]\! g(t) dt,
\end{equation}
\begin{equation}
A_2\!=\!2\!\int_{-\infty}^{\infty}  \! \! g(v^-_2\!-\!v^-_1\!+\!t\!+\!1) \left[\prod_{k=3}^m G(v^-_2\!-\!v^-_k\!+\!t\!+\!1)\right]\! g(t)dt,
\end{equation}
\begin{equation}
B_{(1, j)}\!=\! \int_{-\infty}^{\infty} \! \! g(v^-_1\!-\!v^-_j\!+\!t\!+\!1) G(v^-_1-v^-_2+t+1) \cdot \left[\prod_{\substack{k=3\\k \neq j}}^m \!G(v^-_1\!-v^-_k\!+\!t\!+\!1)\right] g(t)dt,
\end{equation}
\begin{equation}
B_{(2, j)}\!=\! \int_{-\infty}^{\infty} \! \! g(v^-_j\!-\!v^-_1\!+\!t\!+\!1) G(v^-_j\!-\!v^-_2\!+\!t\!+\!1) \cdot \left[\prod_{\substack{k=3\\k \neq j}}^m \!G(v^-_j\!-\!v^-_k\!+\!t\!+\!1)\right] g(t)dt,
\end{equation}
\begin{equation}
B_{(3, j)}\!=\! \int_{-\infty}^{\infty} \! \! g(v^-_j\!-\!v^-_2\!+\!t\!+\!1) G(v^-_j\!-\!v^-_1\!+\!t\!+\!1) \left[\prod_{\substack{k=3\\k \neq j}}^m \! G(v^-_j\! -\! v^-_k\! +\! t\! +\! 1)\right] g(t)dt, 
\end{equation}
\begin{equation}
B_{(4, j)}\!=\! \int_{-\infty}^{\infty} \! \! g(v^-_2\!-\!v^-_j\!+\!t\!+\!1)G(v^-_2\!-\!v^-_1\!+\!t\!+\!1) \left[\prod_{\substack{k=3\\k \neq j}}^m \! G(v^-_2\!-\!v^-_k\!+\!t\!+\!1)\right] g(t)dt.
\end{equation}

We now show that both $A_1 - A_2$ and $B_{(1, j)} - B_{(2, j)} + B_{(3, j)} - B_{(4, j)}$ are non-positive. However, let us first recall some properties of log-concave functions.

\begin{prop}[{\cite[Lemma 1]{an1997log}}]
Consider $g: \mathbb{R} \to \mathbb{R}_+$  and suppose that $\{x : g(x) > 0\} = (a,b)$. Then, $g(x)$ is log-concave if and only if for all $a < x_1 \leq x_2 < b$ and all $\delta \geq 0$ it holds that 
\begin{equation}
g(x_1 + \delta) g(x_2) \geq g(x_1) g(x_2 + \delta).
\end{equation}
\label{prop1:log_concave}
\end{prop}

\begin{prop}[{\cite[Remark 2]{bagnoli2005log}}]
Suppose ${g: \mathbb{R} \to \mathbb{R}_+}$ is a continuously differentiable function and let ${\{x : g(x) > 0\} = (a,b)}$. Then, $g(x)$ is log-concave if and only if $\frac{g'(x)}{g(x)}$ is a non-increasing function of $x$ in $(a,b)$.
\label{prop2:log_concave} 
\end{prop}
\vspace{2mm}
We now prove that $A_1-A_2 \leq 0$. By a change of variable in $A_1$ we let $ v^-_1-v^-_2+t=u$. Then,
\begin{equation}
A_1-A_2 = \int_{-\infty}^{\infty} \prod_{k=3}^m G(u+v^-_2-v^-_k+1) \cdot \left[g(u\!+\! 1) g(u\!+\!v^-_2\!-\!v^-_1 )-g(u) g(u\!+\!v^-_2\!-\!v^-_1\!+\!1 )\right] \, du.
\end{equation}
We now apply Proposition~\ref{prop1:log_concave} to the preceding equation by noting that $u \geq u+v^-_2-v^-_1$ (due to the non-increasing order of the elements in $v^-$), and write 
\begin{equation}
g(u\! +\! 1) g(u+v^-_2-v^-_1 )\! -\! g(u) g(u\! +\! v^-_2\! -\! v^-_1\! +\! 1 )\leq 0.
\end{equation}
Since $\prod_{k=3}^m G(u+v^-_2-v^-_k+1) \geq 0$, we conclude that 
\begin{equation}
A_1 - A_2 \leq 0.
\label{A1-A2}
\end{equation}

Similarly, Proposition~\ref{prop1:log_concave} and Proposition~\ref{prop2:log_concave} can be used to show that $B_{(1,j)}-B_{(2,j)}+B_{(3,j)}-B_{(4,j)} \leq 0$ for all $j=3, \ldots, m$. Therefore, we have verified Schur's condition for $f(v^-)$, and conclude that ${\mathcal{L}(V(x_i') \! \to \! Y_i' \! \mid \! V^-(x_i')=v^-)}$ is Schur-concave.  Finally, by Proposition~\ref{prop:max_min_schur_concave}, the entrywise leakage is maximized by 
\begin{equation}
v^-= v_{max}^- = \left(\frac{L-1}{m}, \ldots, \frac{L-1}{m}\right),
\end{equation}
and is minimized by 
\begin{equation}
v^- = v_{min}^- = (0, \ldots, 0, L-1, 0, \ldots, 0) = (L-1)\, \delta_j,
\end{equation}
for each $j \in \llbracket 1, m\rrbracket$.
\section{Proofs for the leakage with Laplace noise}
\subsection{Proof of Proposition~\ref{prop:data_independent}}\label{ssec:proof_prop2}
Let $N = (N_1, \ldots, N_m)$ be the sequence of i.i.d. Laplace random variables, where $N_j \sim \mathrm{Lap}({\frac{1}{\gamma}})$ for all $j \in \llbracket 1,m \rrbracket$. To find an upper bound on the leakage, we will apply Theorem~\ref{theo:max_individual_leakage} and calculate $\mathcal{L}(V(x_i') \to Y_i' \mid V^-(x_i')=v^-)$ for $v^-= v_{max}^- = \left(\frac{L-1}{m}, \ldots, \frac{L-1}{m}\right)$. We write
\begin{equation}
\mathcal{L}(V(x_i') \to Y_i' \mid V^-(x_i')=v^-_{max}) = \log \sum_{j=1}^m \mathbb{P}(Y_i'=j \mid V(x_i') = v^-_{max} + \delta_j),
\end{equation}
where
\begin{align}
\begin{split}
\mathbb{P}(Y_i'=j \mid V(x_i') = v^-_{max} + \delta_j) &= \mathbb P \{N_j+1> N_1, \ldots, N_j+1> N_m\} \\
&= \int_{-\infty}^{\infty} \left[ \prod_{\substack{l=1\\ l \neq j}}^m \mathbb P \{N_l <(t +1) \} \right]  \cdot \frac{\gamma}{2} \, e^{-\gamma \, \lvert t \rvert} dt, \\
\end{split}
\end{align}
and
\begin{equation}
\mathbb P \{N_l <(t +1) \} = 
\begin{cases}
\frac{1}{2} e^{\gamma (t+1)} & t\leq -1, \\
1-\frac{1}{2} e^{-\gamma (t+1)} & t\geq -1. \\
\end{cases}
\end{equation}
Thus, we have 
\begin{align}
\begin{split}
  \mathbb{P}(Y_i'\!=\!j \!\mid & \! V(x_i') = v^-_{max} \!+\! \delta_j) = \underbrace{\frac{\gamma}{2}\int_{-\infty}^{-1} \! \left[ \frac{1}{2} e^{\gamma(t+1)} \right]^{m-1} \! 
    \! \cdot e^{\gamma t} dt}_{A} \\[1mm]
    &+ \underbrace{\frac{\gamma}{2} \int_{-1}^{0} \left[ 1 - \frac{1}{2} e^{-\gamma(t+1)} \right]^{m-1} \cdot e^{\gamma t} dt}_{B} + \underbrace{\frac{\gamma}{2} \int_{0}^{\infty} \left[ 1 - \frac{1}{2} e^{-\gamma(t+1)} \right]^{m-1} \! \cdot e^{-\gamma t} dt}_{C}. \\[1mm]  
\end{split}
\end{align}
It is straightforward to calculate integrals $A$ and $C$ as
\begin{equation}
A = \frac{2^{-m}}{m} e^{-\gamma} \quad \text{and} \quad C = \frac{1 - \left[ 1 - \frac{1}{2} e^{-\gamma}\right]^m}{m} e^{\gamma}.   
\end{equation}
Integral $B$ can be written as 
\begin{equation}
B = \frac{1}{2} \left(1 - \frac{1}{2} e^{-\gamma} \right)^{m-1} - 2^{-m} e^{-\gamma} - \frac{\gamma (m\!-\!1)}{4} e^{-\gamma} \! \int_{-1}^{0} \! \left(1 - \frac{1}{2} e^{-\gamma (t + 1)} \right)^{m-2} \! \! dt.
\end{equation}
We define 
\begin{align}
\begin{split}
H(m) & \coloneqq \gamma \int_{-1}^0 \left(1 - \frac{1}{2} e^{-\gamma (t + 1)} \right)^m dt \\
&= \gamma \sum_{k=0}^m \binom{m}{k} \left(-\frac{1}{2}\right)^k e^{-\gamma k} \int_{-1}^0 e^{-\gamma kt} dt \\
&= \sum_{k=0}^m \binom{m}{k} \left(-\frac{1}{2}\right)^k \frac{1}{k} \left(1 - e^{-\gamma k} \right). 
\end{split}
\end{align}
Using $\binom{m}{k} = \binom{m-1}{k} + \binom{m-1}{k-1}$ for $m \geq 1$, we get 
\begin{align}
\begin{split}
H(m) &= \! \sum_{k=0}^{m-1} \!  \binom{m-1}{k} \! \left(-\frac{1}{2}\right)^k \! \frac{1}{k} \left(1 \!-\! e^{-\gamma k} \right) + \sum_{k=0}^{m} \binom{m-1}{k-1} \left(-\frac{1}{2}\right)^k \frac{1}{k} \left(1 - e^{-\gamma k} \right) \\ 
&= H(m-1) + \frac{1}{m} \left(2^{-m} - (1 - \frac{1}{2} e^{-\gamma})^m \right),
\end{split}
\end{align}
and $H(0) = \gamma$. Thus, 
\begin{equation}
H(m) = 
\begin{cases} 
\gamma & m=0, \\
\gamma + \displaystyle \sum_{k=1}^m \frac{2^{-k} - (1 - \frac{1}{2} e^{-\gamma})^k}{k} & m \geq 1,
\end{cases}
\end{equation}
Note that $H(m)$ is non-negative and monotonically decreasing in $m$. Since $\sum_{k=1}^\infty \frac{t^k}{k} = \log \frac{1}{1-t}$ for $\abs{t} < 1$, we have $\lim_{m \to \infty} H(m) = 0$. Hence, integral $B$ can be written as 
\begin{equation}
B \!=\! \frac{1}{2} \! \left(1 - \frac{1}{2} e^{-\gamma} \right)^{m-1} \hspace{-4mm} - 2^{-m} e^{-\gamma} - \frac{m-1}{4}  e^{-\gamma} H(m-2).
\end{equation}
Finally, we have 
\begin{equation}
\mathcal{L}(V(x_i') \!  \to \! Y_i' \!  \mid \! V^-(x_i')=v^-) \leq \mathcal{L}(V(x_i') \! \to \! Y_i' \! \mid \! V^-(x_i')=v^-_{max}) = \log (B_1),
\end{equation}
where 
\begin{equation}
B_1 \coloneqq (1-m)\, 2^{-m} e^{-\gamma} + e^{\gamma} \left(1-{(1-\frac{1}{2}e^{-\gamma})}^m\right) + \frac{m}{2} {(1-\frac{1}{2}e^{-\gamma})}^{m-1} - \frac{m(m-1)}{4} e^{-\gamma}  H(m-2).
\end{equation}

\subsection{Proof of Theorem~\ref{theo:ind_leakage}}\label{ssec:proof_theo2}
In order to prove the bound, we will show that ${ k(m) \coloneqq \exp{\Big(\mathcal{L}(V(x_i') \to Y_i' \mid V^-(x_i')=v^-_{max}) [m] \Big)}}$ is concave in $m$ and that 
\begin{equation}
\lim_{m \to \infty} \exp{\Big(\mathcal{L}(V(x_i') \to Y_i' \mid V^-(x_i')=v^-_{max})\Big)} = e^{\gamma}.
\label{eq:limit_m}
\end{equation}
Since $m$ is an integer, we will check the second-order difference of the leakage with respect to $m$. The first-order difference is 
\begin{align}
\Delta k(m) &= k(m+1)- k(m) \\ 
&= (1 - \frac{1}{2} e^{-\gamma})^m - \frac{1}{2} e^{-\gamma} \left(2^{-(m-1)} + m H(m-1) \right), \nonumber
\end{align}
and the second-order difference is 
\begin{align}
\begin{split}
\Delta^2 k(m) &= \Delta k(m+1) - \Delta k(m) \\ 
&= -\frac{1}{2} e^{- \gamma} H(m) \\
&\labelrel\leq{leakage_concave} 0, 
\end{split}
\end{align}
where~\eqref{leakage_concave} follows from the fact that $H(m)$ is non-negative. Thus, we have shown that $\exp\Big(\mathcal{L}(V(x_i') \to Y_i' \mid V^-(x_i')=v^-_{max})\Big)$ is concave in $m$. Furthermore, it is straightforward to verify that~\eqref{eq:limit_m} holds. Hence, we have
\begin{equation}
\mathcal{L}(V(x_i') \to Y_i' \mid V^-(x_i')=v^-_{max}) \leq \gamma.
\end{equation}
Finally, we get 
\begin{align}
\begin{split}
\mathcal L (D^* \to & Y_i' \mid D^- = d^-) = {\mathcal L}(D \to Y_i' \mid D^-=d^-) \\
& \leq \mathcal{L}(V(x_i') \to Y_i' \mid V^-(x_i')=v^-) \\
& \leq \mathcal{L}(V(x_i') \to Y_i' \mid V^-(x_i')=v^-_{max})\\
&\leq \gamma.
\end{split}
\end{align} 

\subsection{Proof of Proposition~\ref{prop:data_dependent}}\label{ssec:proof_prop_data_dependent}
Similarly to the proof of Proposition~\ref{prop:data_independent}, we can write 
\begin{equation}
\mathcal{L}(V(x_i') \to Y_i' \mid V^-(x_i')=v^-)= \log \sum_{j=1}^m \mathbb{P}(Y_i'=j \mid V(x_i') = v^- + \delta_j), \hspace{0.5cm}
\end{equation}
where 
\small
\begin{equation}
    \mathbb{P}(Y_i'=j \mid V(x_i') = v^- + \delta_j) = \mathbb P \{N_j+v^-_j+1 \! > \! N_1+v^-_1, \ldots, N_j+v^-_j+1 \!> \! N_m+v^-_m\}.
\end{equation}
\normalsize
For $1 \leq j \leq r$, we have 
\begin{align}
\begin{split}
    \mathbb{P}(Y_i'=j \mid V(x_i') = \, v^- + \delta_j) &= \mathbb{P}(Y_i'=1 \mid V(x_i') = v^- + \delta_1) \\
    &\leq \mathbb{P}\{N_1+v^-_1 + 1 > N_2 + v^-_2\} \\ 
    &= \mathbb{P}\{N_2- N_1 < v^-_1 + 1 - v^-_2 \},
\end{split}
\end{align}
and for $r+1 \leq j \leq m$, we have 
\begin{align}
\begin{split}
    \mathbb{P}(Y_i'=j \mid V(x_i') = \, v^- + \delta_j) &\leq \mathbb{P}\{N_j+v^-_j + 1 > N_1 + v^-_1\} \\ 
    &= \mathbb{P}\{N_1- N_j < v^-_j + 1 - v^-_1 \}. 
\end{split}
\end{align}
It is straightforward to see that the random variable described as the difference of two $\mathrm{Lap}(\frac{1}{\gamma})$ random variables has the following CDF: 
\small
\begin{equation}
    \mathbb{P}\{N_1 - N_2 \leq x\} = \begin{cases}
    \frac{1}{4} \exp(\gamma x) (2 - \gamma x) & x \leq 0, \\
    1 - \frac{1}{4} \exp(-\gamma x) (2 + \gamma x) & x \geq 0.
    \end{cases}
\end{equation}
\normalsize
Then, by noting that $v^-_1 + 1 - v^-_2 > 0$ and $v^-_j + 1 - v^-_1 \leq 0$ for $r+1 \leq j \leq m$, we get 
\begin{equation}
    \mathcal{L}(V(x_i') \to Y_i' \mid V^-(x_i')=v^-) \leq \log (B_2),
\end{equation}
where 
\begin{equation}
    B_2 \coloneqq r \left(1 - \frac{2 + \gamma (v^-_1 + 1 - v^-_2)}{4 \exp\left(\gamma(v^-_1 + 1 - v^-_2)\right)} \right) + \sum_{j=r+1}^m \frac{2 + \gamma (v^-_1 - 1 - v^-_j)}{4 \exp \left( \gamma (v^-_1 - 1 - v^-_j)\right)}. \hspace{0.5cm}
    \IEEEQED
\end{equation}


\ifCLASSOPTIONcaptionsoff
  \newpage
\fi



%
\bibliographystyle{IEEEtran}
\bibliography{mybib}

\end{document}